\documentclass[12pt,thmsa]{article}
\usepackage{sw20lart}

\input tcilatex

\begin{document}

\author{I. Radinschi\thanks{%
iradinsc@phys.tuiasi.ro} \and Department of Physics, ''Gh. Asachi''
Technical University, \and Iasi, 6600, Romania}
\title{Energy Distribution of a Black Hole Solution in Heterotic String Theory }
\date{The Date }
\maketitle

\begin{abstract}
We calculate the energy distribution of a charged black hole solution in
heterotic string theory in the M\o ller prescription.


PACS: 04. 20 Dw, 04. 70. Bw
\end{abstract}

\section{INTRODUCTION}

One of the most interesting problem of relativity is the energy and momentum
localization. The different attempts at constructing an energy-momentum
density don't give a generally accepted expression. However, there are
various energy-momentum complexes including those of Einstein [1]-[2],
Landau and Lifshitz [3], Papapetrou [4], Bergmann [5], Weinberg [6] and M\o
ller [7]. Cooperstock [8] gave his opinion that the energy and momentum are
confined to the regions of non-vanishing energy-momentum tensor of the
matter and all non-gravitational fields. Although, the energy-momentum
complexes are coordinate dependent they can give a reasonable result. To get
meaningful results for the energy distribution using the energy-momentum
complexes of Einstein, Landau and Lifshitz, Papapetrou, Bergmann and
Weinberg the calculations are carried out in Cartesian coordinates. Some
interesting results recently obtained sustain this conclusion [9]-[14].

The M\o ller energy-momentum complex [7] no needs to carry out calculations
in Cartesian coordinates so we can calculate in any coordinate system. Some
results recently obtained [15]-[18] sustain that the M\o ller
energy-momentum complex is a good tool for obtaining the energy distribution
in a given space-time. Also, in his recent paper, Lessner [19] gave his
opinion that the M\o ller definition is a powerful concept of energy and
momentum in general relativity. Also, Chang, Nester and Chen [20] showed
that the energy-momentum complexes are actually quasilocal and legitimate
expression for the energy-momentum. They concluded that there exist a direct
relationship between energy-momentum complexes and quasilocal expressions
because every energy-momentum complexes is associated with a legitimate
Hamiltonian boundary term.

Sen metric [24] is new and is becoming very important. So, we find to be of
interest to study its energy distribution. The aim of this paper is to
calculate the energy distribution of this charged black hole solution in
heterotic string theory in the M\o ller prescription. We use geometrized
units ($G=1,c=1$) and follow the convention that Latin indices run from 0 to
3.

\section{ENERGY\ IN THE M\O LLER PRESCRIPTION}

The low-energy effective field theory describing string theory contains
black hole (or, more generally, black p-brane) solutions which can have
properties which are qualitatively different from those that appear in
ordinary Einstein gravity [21].

Rotating charge-neutral black hole solutions can be constructed in string
theory, and are identical to the Kerr solution [22] of ordinary Einstein
gravity with the dilaton taking a constant value. Rotating charged black
hole solutions in these theories have been analyzed [23] in the limit of
small angular momentum.

The solution constructed by A. Sen [24] is an exact classical solution in
the low-energy effective field theory, which describes a black hole carrying
a finite amount of charge and angular momentum. The method he used is the
twisting procedure (see references [4]-[8] in [24]) that generates
inequivalent classical solutions starting from a given classical solution of
string theory. He generated the rotating charged black hole solution by
starting from a rotating black hole solution carrying no charge, like the
Kerr solution [22].

For the black hole solution obtained by A. Sen we consider the case when the
angular momentum is zero. We have two cases.

In the first case, the metric (obtained from eq. (11) in [24]) is given by

\begin{equation}
ds^{2}=\frac{(r^{2}-2\,m\,r)\,r^{2}}{(r^{2}+2\,m\,r\,\sinh ^{2}(\,\frac{a}{2}%
))^{2}}\,\,dt^{2}-\frac{r^{2}}{r^{2}-2\,m\,r}\,\,dr^{2}-r^{2}\,d\theta
^{2}-r^{2}\,\sin ^{2}\,\theta \,d\varphi ^{2},  \tag{1}  \label{1}
\end{equation}

where $a$ is an arbitrary number. The metric given by (1) describes a black
hole solution with mass $M$ and charge $Q$ given by $M=(\frac m2)\,(1+\cosh
\,a)$ and, respectively, $Q=(\frac m{\sqrt{2}})\,\sinh \,a$.

In the second case, the Einstein metric ($ds_{E}^{2}=e^{-\Phi ^{^{\prime
}}}\,ds^{2}$, with $\Phi ^{^{\prime }}$ given by eq. (12) in [24] and
obtained from eq. (16) in [24]) is given by

\begin{eqnarray}
ds_{E}^{2} &=&\frac{(r^{2}-2\,m\,r)}{(r^{2}+2\,m\,r\,\sinh ^{2}(\,\frac{a}{2}%
))}\,\,dt^{2}-\frac{r^{2}+2\,m\,r\,\sinh ^{2}\,(\frac{a}{2})}{r^{2}-2\,m\,r}%
\,\,dr^{2}-  \TCItag{2} \\
&&-(r^{2}\,+2\,m\,r\,\sinh ^{2}\,(\frac{a}{2}))\,d\theta ^{2}-  \nonumber \\
&&-(r^{2}+2\,m\,r\,\sinh ^{2}\,(\frac{a}{2}))\,\sin ^{2}\,\theta
\,\,d\varphi ^{2}.  \nonumber
\end{eqnarray}

The M\o ller energy-momentum complex $V_{i}^{\;k}$ [7] is given by

\begin{equation}
V_{i}^{\;k}={\frac{1}{8\,\pi }\,}\chi _{i\;\;\;,l}^{\;kl},  \tag{3}
\label{3}
\end{equation}
where

\begin{equation}
\chi _{i}^{\;kl}=-\chi _{i}^{\;lk}=\sqrt{-g}\,\left( {\frac{\partial g_{in}}{%
\partial x^{m}}}-{\frac{\partial g_{im}}{\partial x^{n}}}\right)
\,g^{km}\,g^{nl}.  \tag{4}  \label{4}
\end{equation}

Also, $V_{i}^{\;k}$ satisfies the local conservations laws

\begin{equation}
{\frac{\partial V_{i}^{\;k}}{\partial x^{k}}}=0.  \tag{5}  \label{5}
\end{equation}

$V_{0}^{\;\,0}$ is the energy density and $V_{\alpha }^{\;\,0}$ are the
momentum density components.

The energy is given by

\begin{equation}
E=\int \int \int V_{0}^{\,\,\,0}dx^{1}dx^{2}dx^{3}={\frac{1}{8\,\pi }}\int
\int \int {\frac{\partial \chi _{0}^{\,\,0l}}{\partial x^{l}}\,}%
dx^{1}\,dx^{2}\,dx^{3}.  \tag{6}
\end{equation}

For the metric given by (1) the $\chi _0^{\;01}$ component is given by

\begin{equation}
\chi _{0}^{\;01}=-2\,\frac{m\,\sin \,\theta \,r\,(-2\,r\,\sinh ^{2}\,(\frac{a%
}{2})-r+2\,m\,\sinh ^{2}\,(\frac{a}{2}))}{(r+2\,m\,r\,\sinh ^{2}\,(\frac{a}{2%
}))^{2}}.  \tag{7}  \label{7}
\end{equation}

Using (6) and (7), after some calculations and applying the Gauss theorem we
obtain the energy distribution

\begin{equation}
E=m\,(1+\frac{a^{2}}{2}+\frac{a^{4}}{24}+\frac{a^{6}}{720}+O(a^{7})). 
\tag{8}  \label{8}
\end{equation}

The graphic representation for the energy distribution is given in the Fig. 1

From (8) we observe that the energy distribution depends on $a$. In the case 
$a=0$ we obtain

\begin{equation}
E=m.  \tag{9}  \label{9}
\end{equation}

For the metric given by (2) the $\chi _0^{\;01}$ component is given by

\begin{equation}
\chi _{0}^{\;01}=2\,\frac{m\,\sin \,\theta \,r\,(1+\,\sinh ^{2}\,(\frac{a}{2}%
))}{r+2\,m\,\,\sinh ^{2}\,(\frac{a}{2})}.  \tag{10}  \label{10}
\end{equation}

Using (10) and (6), after some calculations and applying the Gauss theorem
we obtain the energy distribution

\begin{equation}
E=m\,(1+\frac{a^{2}}{4}+\frac{a^{4}}{48}+\frac{a^{6}}{1440}+O(a^{7})). 
\tag{11}  \label{11}
\end{equation}

For the energy distribution we have the graphic representation from the Fig.
2

Also, in this case if we have $a=0$ we obtain

\begin{equation}
E=m.  \tag{12}  \label{12}
\end{equation}

\section{DISCUSSION}

Bondi [25] gave his opinion that a nonlocalizable form of energy is not
admissible in relativity. As we pointed out [9]-[18], there are many
important results recently obtained which sustain the view point of Bondi.

For the metrics given by (1) and (2) we obtained the expressions for the
energy distribution in the M\o ller prescription given by (8) and (11). The
first term is the same in both (8) and (11) and the second, the third and
the fourth terms in (8) are twice than their values in (11). The energy
distribution depends on the arbitrary number $a$. Also, because the
dependence of $m$ and $\sinh \,(\frac{a}{2})$ on the independent physical
parameters $M$ and $Q$ (see eqs. (19) in [24]), the energy distribution
depends on the mass $M$ and charge $Q$ of the black hole. In the case $a=0$
we obtain that $E=m$. These results sustain that the M\o ller
energy-momentum complex is a powerful tool for obtaining the energy
distribution in a given space-time.

Sen \ metric represents black hole as well as naked singularities depending
on the ratio of the mass to the electric charge parameters. Recently,
Virbhadra and a very reputed general relativist Ellis, pioneered
gravitational lensing by very strong fields of black holes as well as naked
singularities[26] in order to test the well-known Cosmic Censorship
Hypothesis of Roger Penrose. I suggest that these studies be repeated with
the Sen metric. This will help testing the low energy string theory in
comparison with the Einstein-Maxwell theory and will also show the role of
the total mass parameter (calculated in this paper) on gravitational lensing
phenomena.


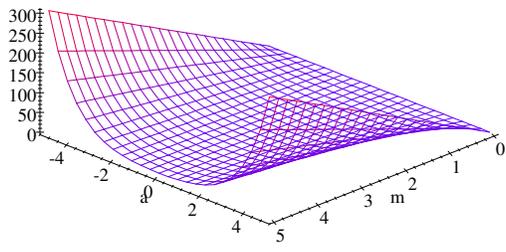
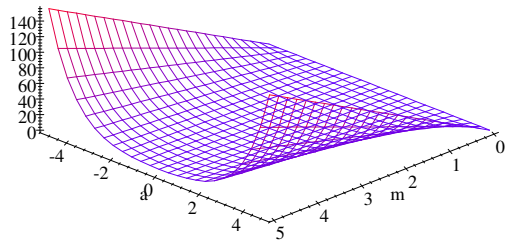
\end{document}